\documentclass[hyper]{JHEP} 

\usepackage{epsfig}

\newcommand\fverb{\setbox\pippobox=\hbox\bgroup\verb}
\newcommand\fverbdo{\egroup\medskip\noindent%
			\fbox{\unhbox\pippobox}\ }
\newcommand\fverbit{\egroup\item[\fbox{\unhbox\pippobox}]}
\newbox\pippobox
\title{Marginal Deformations In the Open Bosonic String
Field Theory for N
D0-branes}
\author{ J. Kluso\v{n}
\footnote{On leave from Masaryk University, Brno}\\
Institute of Theoretical Physics, University of Stockholm, SCFAB\\
SE- 106 91 Stockholm, Sweden \\
and \\
Institutionen f\"or teoretisk fysik\\
BOX 803, SE- 751 08 
Uppsala, Sweden \\
E-mail: \email{josef.kluson@teorfys.uu.se}}

\preprint{\hepth{0203089}}

\abstract{In this short note we give an
example of the exact solution of the open
bosonic string field theory defined on 
the background of $N$
coincided D0-branes. This  solution
leads to the change of the original background
to the background where D0-branes are
localised in general positions.}

\def\tr{\mathrm{Tr}}

\def\ket #1{\left|#1\right>}

\begin{document}
\section{Introduction}\label{first}
In our recent paper \cite{KlusonESFT2} we gave
examples of two exact solutions of the 
open bosonic string field theory 
\cite{WittenSFT} that correspond
to the marginal deformation of the original
background configuration of D0-brane sitting
in the origin.  These two
 examples served as a demonstration of
the  efficiency of our method developed
in \cite{KlusonESFT1}. We 
have seen that the solutions given in
\cite{KlusonESFT2} did not change 
the form of the BRST operator and hence the
spectrum of the string states 
 around the new background
is the same. As was argued in that paper
 this result is a
trivial consequence of the homogeneity and isotropy
of the space where D0-brane lives. 
We have also argued  that in
the case  of $N$ D0-branes 
the situation will not
be so simple. The analysis of this configuration
 will be performed
in this paper where we find exact solution of 
the open bosonic string field theory 
that corresponds to the marginal deformation of the original
configuration. We will see that when we 
perform   expansion around the new background
configuration we obtain the string field theory
action for the fluctuation field that has the same
form as the original one however the new BRST operator
is matrix valued as a consequence of the nonabelian
nature of the initial configuration. This result
supports the proposal given in
\cite{KlusonMSFT}, however this claim certainly deserves
much more detailed study and we will postpone it for
the future. 
Then we will argue, following
our general discussion given in \cite{KlusonESFT1}, that
we can perform the field redefinition that
maps the new  BRST operator to the original one and
the  fluctuation field that arises from the 
expansion of the string field around 
the classical solution of the string field theory
 equation of
motion to  
 the fluctuation field
with nonzero winding number. More precisely,
we  will see that
the  state of the string joining i-Th and
j-Th D0-branes has the winding number proportional
to the distance of these two D0-branes.

Since this paper is closely related to
 \cite{KlusonESFT1,KlusonESFT2}
we will not discuss the general construction presented
in \cite{KlusonESFT1} and the abelian
solution given   in \cite{KlusonESFT2},
where more details  and more detailed list of
references can be found. However it
is a pleasant duty for us mention some of the
papers 
\cite{Takahashi1,Takahashi2,Takahashi3,SenM1,Gaiotto:2002uk}
that discuss related problems from different
point of view.

In the conclusion we outline our results and suggest
possible extension of our work.

\section{Open bosonic string field theory
for N D0-branes}
In this section we present nonabelian generalization
of our result given in \cite{KlusonESFT2}.
For more details and motivation, see that paper.

 Let us consider system of coincident
$N$ D0-branes localized in the origin of
the target space. As was argued in 
\cite{WittenSFT} such a system
is described with the standard open
bosonic
string field theory where now
string field will carry Chan-Paton
(CP) indexes
 so that we can write
\begin{equation}\label{Naction}
S=-\sum_{i,j,k}\int\frac{1}{2}\Phi_{ij} 
\star Q(\Phi_{ji})+
\frac{1}{3}\Phi_{ij} \star\Phi_{jk}
 \star  \Phi_{ki} \ ,
\end{equation}
where  $\Phi_{ij}$ is a  ghost number one string field and
the summation index goes from $1$ to $N$ and
where $Q$ is the BRST operator of open string
that obeys Dirichlet boundary conditions
in $x^I, I=1,\dots ,25$ directions
(For more details about string
field theory, see very good reviews
\cite{Ohmori,Desmet,BelovR}.).
From (\ref{Naction}) we obtain the
  equation of
 motion
\begin{equation}\label{eqmnonabelian}
Q\Phi_{ij}+\sum_k\Phi_{ik}\star\Phi_{kj}=0 \  .
\end{equation}
The generalization of the solution given
in \cite{KlusonESFT2} to the nonabelian
case  is straightforward 
\begin{equation}
\Phi_{0ij}=\sum_k(
e^{-\Psi_0})_{ik}\star Q(e^{\Psi_0})_{kj}
 \ .
\end{equation}
It is easy  to see that $\Phi_{0ij}$
with any ghost number zero field
$\Psi_{0ij}$ is   solution of the equation of motion
(\ref{eqmnonabelian}) since we have
\begin{equation}
Q(\Phi_0)_{ij}=Q\left(\sum_k(e^{-\Psi_0})_{ik}
\star Q(e^{\Psi_0})_{kj}\right)=
-\sum_{k,l,m}(e^{-\Psi_0})_{ik}\star Q(e^{\Psi_0})_{kl}
\star (e^{-\Psi_0})_{lm}
\star Q(e^{-\Psi_0})_{mj}
\end{equation}
and
\begin{equation}
\sum_k(\Phi_0)_{ik}\star (\Phi_0)_{kj}
=\sum_{k,m,l}(e^{-\Psi_0})_{ik}
\star Q(e^{\Psi_0})_{kl}
\star (e^{-\Psi_0})_{lm}\star Q(e^{\Psi_0})_{mj} \ .
\end{equation}
As in   case of single D0-brane
\cite{KlusonESFT2}
  we propose that  the field $\Psi_{0ij}$
has  the form
\begin{equation}\label{solnab}
(\Psi_0)_{ij}=\hat{K}_L(\hat{I},F)_{ij} \ ,
\end{equation}
where we define the generalization of
the identity field of the string star algebra
 to the nonabelian 
case  as 
\begin{equation}
\hat{I}_{ij}=\mathcal{I}\otimes \delta_{ij} \ .
\end{equation}
This field has the basic  property of the identity
element of the string field algebra
\begin{equation}
\sum_k\Psi_{ik}\star \hat{I}_{kj}=
\Psi_{ij}\star \mathcal{I}=\Psi_{ij} 
\ , \sum_k \hat{I}_{ik}\star \hat{I}_{kj}=
\hat{I}_{ij} \ .
\end{equation}
In (\ref{solnab}) we have also introduced the ghost
number zero operator $\hat{K}$.
 Following \cite{KlusonESFT1} we will study the 
fluctuation around the solution $\Phi_{0ij}$. 
We write general string field as follows
\begin{equation}
\Phi_{ij}=\Phi_{0ij}+\Psi_{ij} \ 
\end{equation}
and insert  it into (\ref{Naction}).
According to the well known procedure
(For recent discussion, see \cite{SenV5}.) we obtain
the action for the field $\Psi_{ij}$ which 
has the same form as the original one (\ref{Naction})
however with the new BRST operator
\begin{equation}\label{QtildeG}
\tilde{Q}(X)_{ij}=\sum_{k}
\delta_{ik}Q(X)_{kj}+\Phi_{0ik}\star X_{kj}-
(-1)^{|X|}X_{ik}\star \Phi_{0kj} \ .
\end{equation}
According to \cite{KlusonESFT1} this new BRST operator
$\tilde{Q}$ has a form
\begin{equation}\label{QtildeG2}
\tilde{Q}(X)_{ij}=
e^{-\hat{K}}\left(Q\left
(e^{\hat{K}}(X)\right)\right)_{ij} \ .
\end{equation}
In order the string field theory for fluctuation field
$\Psi_{ij}$ to be well defined we
 must show that the  BRST operator
(\ref{QtildeG2}) obeys all axioms of
the open bosonic
string field theory \cite{Ohmori,Desmet,BelovR}.

Let us start with the proof of the nilpotence
of (\ref{Qtilde})\footnote{We do not write
summation index for the summation
over $k,l$ in following expressions.}
\begin{eqnarray}
\tilde{Q}^2(X_{ij})=\tilde{Q}\left(
Q(X)_{ij}+\Phi_{0ik}\star X_{kj}-(-1)^{|X|}
X_{ik}\star \Phi_{0kj}\right)=\nonumber \\
=Q\left(
Q(X)_{ij}+\Phi_{0ik}\star X_{kj}-(-1)^{|X|}
X_{ik}\star \Phi_{0kj}\right)+\nonumber \\
+\Phi_{0ik}\star \left(
Q(X)_{kj}+\Phi_{0kl}\star X_{lj}-(-1)^{|X|}
X_{kl}\star \Phi_{0lj}\right) - \nonumber \\
-(-1)^{|X|+1}\left(
Q(X)_{ik}+\Phi_{0ik}\star X_{kl}-(-1)^{|X|}
X_{ik}\star \Phi_{0kl}\right)\star \Phi_{0lj}=
\nonumber \\
=(Q(\Phi_{0ik})+\Phi_{0il}\star \Phi_{0lk})
\star X_{kj}-(-1)^{2|X|}X_{ik}\star (
Q(\Phi_{0kj})+\Phi_{0kl}\star \Phi_{0lj})=0
\nonumber \\
\end{eqnarray}
since $\Phi_{0ij}$ obeys (\ref{eqmnonabelian}).
Derivation property can be seen very easily
since we have
\begin{eqnarray}
\tilde{Q}(X\star Y)_{ij}=
Q(X)_{ik}\star Y_{kj}+(-1)^{|X|}X_{ik}
\star Q(Y)_{kj}+\Phi_{0ik}\star X_{kl}
\star Y_{lj}-\nonumber \\
-(-1)^{|X|+|Y|}X_{ik}\star
Y_{kl}\star \Phi_{0lj}
=(Q(X)_{il}+\Phi_{0ik}\star X_{kl}-
(-1)^{|X|}X_{ik}\star \Phi_{0kl})\star
Y_{lj}+\nonumber \\
+(-1)^{|X|}X_{ik}\star
(Q(Y)_{kj}+\Phi_{0kl}\star Y_{lj}-
(-1)^{|Y|}Y_{kl}\star \Phi_{0lj})=\nonumber \\
=
\tilde{Q}(X)_{ik}\star Y_{kj}+
(-1)^{|X|} X_{ik}\star \tilde{Q}(Y)_{kj} \ .
\nonumber \\
\end{eqnarray}
Finally we have
\begin{eqnarray}
\sum_{i,j}\int \tilde{Q}(X)_{ij}=
\sum_{i,j}\ \int Q(X)_{ij}+\int
\Phi_{0ij}\star X_{ji}-
(-1)^{|X|}\int X_{ij}\star \Phi_{0ji}
=\nonumber \\
=\sum_{i,j}\int
\Phi_{0ij}\star X_{ji}-
(-1)^{|X|}(-1)^{|X||
\Phi_0|}\int  \Phi_{0ji}\star X_{ij}=0 \ ,
\nonumber \\
\end{eqnarray}
using $|\Phi_0|=1$. As a consequence the new BRST
operator $\tilde{Q}$ obeys all axioms of the open
bosonic string field theory even in the case when
it is matrix valued.
However it is desirable to
have the BRST operator in the same form as the original
one which means that will be diagonal in 
the CP indexes. In order to obtain such a form
we should perform field redefinition according
to \cite{KlusonESFT1}.
 We will see that with the solution
given in (\ref{solnab}) such a redefinition is
very simple.

From (\ref{QtildeG2}) we immediately see that it
is natural to perform field redefinition
\begin{equation}\label{redG}
\Psi_{ij}=e^{-\hat{K}}(\tilde{\Psi})_{ij} \ 
\end{equation}
that maps the new BRST operator (\ref{QtildeG2})
to the original one. It is important to stress
that $\hat{K}$ is not arbitrary ghost number zero
operator, while according to \cite{SenV5} 
$\hat{K}$ must obey following rules 
\begin{eqnarray}\label{Krules}
\hat{K}(\sum_l X_{il}\star Y_{lj})=
\sum_l\hat{K}(X)_{il}\star Y_{lj}+
\sum_l X_{il}\star \hat{K}(Y)_{lj}
 \  , \nonumber \\
\sum_{i,j} \int \hat{K}(X_{ij}\star Y_{ji})=0 \ .
\nonumber \\
\end{eqnarray}
If we presume that $\hat{K}$ obeys (\ref{Krules})
it is a simple task to show that  
 the field redefinition
(\ref{redG}) leads to the action for 
the field $\tilde{\Psi}_{ij}$  that
has the same form as (\ref{Naction}) and
where the BRST operator is the same as the original
one $Q$. To prove this,  let us start with the quadratic term
\begin{eqnarray}\label{kvad}
\frac{1}{2}\sum_{i,j}\int \Psi_{ij}
\star \tilde{Q}(\Psi_{ji})=\frac{1}{2}
\sum_{i,j}\int e^{-\hat{K}}(\tilde{\Psi})_{ij}
\star\tilde{Q}(e^{-\hat{K}}(\tilde{\Psi}))_{ji}=
\nonumber \\
=\frac{1}{2}\sum_{i,j}\int \tilde{\Psi}_{ij}
\star e^{\hat{K}}(\tilde{Q}
(e^{-\hat{K}}(\tilde{\Psi})))_{ji}
=\frac{1}{2}\sum_{i,j}\int \tilde{\Psi}_{ij}
\star {Q}(\tilde{\Psi})_{ji} \ , \nonumber \\
\end{eqnarray}
where we have used
\begin{eqnarray}
\sum_{i,j}\int e^{-\hat{K}}(X)_{ij}\star
Y_{ji}=\sum_{i,j}\int X_{ij}\star Y_{ji}-
\hat{K}(X)_{ij}\star Y_{ji}+\frac{1}{2}
\hat{K}^2(X)_{ij}\star Y_{ji}+\dots=\nonumber \\
=\sum_{i,j}\int X_{ij}\star Y_{ji}+
X_{ij}\star \hat{K}(Y)_{ji}-
\frac{1}{2}\hat{K}(X)_{ij}\star \hat{K}(Y)_{ji}
+\dots =\nonumber \\
=\sum_{i,j}\int X_{ij}\star Y_{ji}+
X_{ij}\star \hat{K}(Y)_{ji}+
\frac{1}{2}X_{ij}\star \hat{K}^2(Y)_{ji}
+\dots =\sum_{i,j}\int
X_{ij}\star e^{\hat{K}}(Y)_{ji} \ ,\nonumber \\
\end{eqnarray}
and 
\begin{equation}
e^{\hat{K}}(X\star Y)_{ij}=
\sum_k e^{\hat{K}}(
X)_{ik}\star e^{\hat{K}}(Y)_{kj} \ 
\end{equation}
that follows from  (\ref{Krules}). 
Now it is a simple task to show the invariance
of the cubic term in (\ref{Naction})
 under field redefinition since
\begin{eqnarray}\label{vertex}
\sum_{i,j,k}\int \Psi_{ij}\star \Psi_{jk}
\star \Psi_{ki}=\sum_{i,j,k}
\int e^{-\hat{K}}(\tilde{\Psi})_{ij}\star
e^{-\hat{K}}(\tilde{\Psi})_{jk}\star
e^{-\hat{K}}(\tilde{\Psi})_{ki}=\nonumber \\
=\sum_{i,j}\int
\tilde{\Psi}_{ij}\star e^{\hat{K}}
\left(e^{-\hat{K}}(\tilde{\Psi}\star \tilde{\Psi})
\right)_{ji}=\sum_{i,j,k}\int
\tilde{\Psi}_{ij}\star \tilde{\Psi}_{jk}
\star \tilde{\Psi}_{ki} \ . \nonumber \\
\end{eqnarray}
From (\ref{kvad}) and (\ref{vertex}) we see
that the action for the new fluctuation field
$\tilde{\Psi}$ has the same form as the original
one (\ref{Naction}) which we wanted to prove. 

Now from this general discussion we turn to the
nonabelian extension  of our result given in 
\cite{KlusonESFT2}.
For that reason we must determine the form
of the operator $\hat{K}$ and the meaning of
the function $F_{ij}^I(\sigma)$ that appears in this operator.
In fact, we introduce this function into
$\hat{K}$ according to the  analysis performed in
\cite{Takahashi1,Takahashi3} in order to eliminate
presence of the delta functions 
and also 
to eliminate contribution from the midpoint of
the string.
We demand that this function   must obey these properties
\begin{equation}
F^I_{ij}(0)=F^I_{ij}(\pi)=\frac{1}{\sqrt{2\pi
\alpha'}}Y^I_{ij}, 
F^I_{ij}\left(\frac{\pi}{2}\right)=0 \ .
\end{equation}
where $Y^I_{ij}
, \ i, j=1,\dots, N, I=1,\dots, 25$ are matrices that
describe the marginal deformation of the configuration
of $N$ D0-branes originally sitting in the origin.
We also demand that these matrices commute among themselves
so that can be put into the form
\begin{equation}
Y^I_{ij}=y^I_{i} \delta_{ij}  \ ,
\end{equation}
where $y^I_i$ parametrises shift of the position of the
i-th D0-brane in the I-th direction. 

 Let us start with an
 infinitesimal form of the solution
 (\ref{solnab}) $\Phi_{0ij}$ where we presume that
the deformation parameters in $Y^I_{ij}$ are
small. Then we obtain
\begin{equation}\label{small}
\Phi_{0ij}=Q(\hat{K}_L(\hat{I},F))_{ij} \ .
\end{equation}
Let us define operator $\hat{K}$ as follows
\footnote{In the following we will not write its dependence
of the function $F^I_{ij}$.}
\begin{eqnarray}\label{K}
\hat{K}(X)_{ij}=\sum_k \left( D_{ik}\star X_{kj}-
X_{ik}\star D_{kj}\right)
\ ,
D_{ik}=P_I^L(\mathcal{I}F^I)_{ik}
 \ , \nonumber \\
P_I^L(F^I)_{ij}=\frac{i}{2\pi\alpha'}
\int_0^{\pi/2}d\sigma\eta_{IJ}\dot{X}^J
F^I(\sigma)_{ij} \ ,
P_I^R(F^I)=\frac{i}{2\pi\alpha'}
\int_{\pi/2}^{\pi}d\sigma\eta_{IJ}\dot{X}^J
F^I(\sigma)_{ij} \ , \nonumber \\
\end{eqnarray}
where in the definition of $D_{ij}$ the summation over
$I$ is implicitly presumed. 
From previous expressions it is also clear the meaning of
the symbols $L,R$ that will appear in various operators.
It is clear that 
$\hat{K}$  is nonabelian
generalization of the operator $K$ given
 in \cite{KlusonESFT2} for  system of
$N$ D0-branes. We also see that $\hat{K}$ annihilates 
 identity 
field $\hat{I}=\mathcal{I}\otimes \delta_{ij}$ since we have
\begin{equation}
\hat{K}(\hat{I})_{ij}=\sum_k 
D_{ik}\star \delta_{kj}\mathcal{I}-\delta_{ik}\mathcal{I}
\star D_{kj}=D_{ij}-D_{ij}=0
 \ .
\end{equation}
Then we  define
\begin{equation}
\hat{K}_L(\hat{I})_{ij}=\sum_k D_{ik}\star
\delta_{kj} \mathcal{I}=D_{ij}
 \ .
\end{equation}
We can
easily
show that  $\hat{K}$ obeys 
all axioms given in (\ref{Krules}) 
since we have
\begin{eqnarray}
\hat{K}(\sum_l X_{il}\star Y_{lj})=
\sum_{k,l} D_{ik}\star X_{kl}\star Y_{lj}-
X_{ik}\star Y_{kl}\star D_{lj}=
\nonumber \\
=\sum_{k,l}
D_{ik}\star X_{kl}\star Y_{lj}-
X_{ik}\star D_{kl}\star Y_{lj}+\nonumber \\
+ X_{ik}\star D_{kl}
\star Y_{lj}-X_{ik}\star Y_{kl}
\star D_{lj}=\nonumber \\
=\sum_k \hat{K}(X)_{ik}\star Y_{kj}+
X_{ik}\star \hat{K}(Y)_{kj} \  \nonumber \\
\end{eqnarray}
and
\begin{eqnarray}
\tr\int \hat{K}(\Phi)=
\sum_{i,j}\left(
\int D_{ij}\star \Phi_{ji}-\int \Phi_{ij}
\star D_{ji}\right)=\nonumber \\
=\sum_{i,j}\left(\int D_{ij}\star \Phi_{ji}-
(-1)^{|D||\Phi|}\int D_{ij}\star \Phi_{ji}\right)
=0\nonumber \\
\end{eqnarray}
since the ghost number of 
 $D$ is equal to zero $|D|=0$.

Now we  return to (\ref{small}) that
can be written as \cite{KlusonESFT1,KlusonESFT2}
\begin{equation}\label{phis}
\Phi_{0ij}=Q(\hat{K}_L(\hat{I}))_{ij}=
Q(P_I^L(\mathcal{I}F^I))_{ij}=[Q,P_I(F^I)]^L(\mathcal{I})_{ij}
\end{equation}
and hence the new  BRST operator is
\begin{eqnarray}\label{Qtilde}
\tilde{Q}(X)_{ij}=\sum_{k}
\delta_{ik}Q(X)_{kj}+\Phi_{0ik}\star X_{kj}-
(-1)^{|X|}X_{ik}\star \Phi_{0kj}=
\nonumber \\
=\sum_k
\delta_{ik}Q(X)_{kj}-\tilde{D}^L
(X)_{ij}-\tilde{D}^R(X)_{ij}
\nonumber \\
\end{eqnarray}
using
\begin{eqnarray}
\sum_k\Phi_{0ik}\star X_{kj}=
\sum_k [Q,P_I(F^I_{ik})]^L(\mathcal{I})\star (X_{kj})=
-\sum_k \mathcal{I}\star [Q,P_I(F^I_{ik})]^R
(X_{kj}) \ , \nonumber \\
-\sum_k(-1)^{|X|}X_{ik}\star \Phi_{0kj}=
-(-1)^{|X|}\sum_k X_{ik}
\star [Q,P_I(F^I_{kj})]^L(\mathcal{I})=
\nonumber \\=
(-1)^{|X|}\sum_k X_{ik}\star 
[Q,P_I(F^I_{kj})]^R(\mathcal{I})=-
\sum_k [Q,P_I(F^I_{ik})]^L(X_{kj}) 
\nonumber \\
\end{eqnarray}
and we also  defined 
\begin{equation}
\tilde{D}^L(X)_{ij}=
\sum_k [Q,P_I(F^I_{ik})]^L(X_{kj}), \ 
\tilde{D}^R(X)_{ij}=
\sum_k [Q,P_I(F^I_{ik})]^R(X_{kj}) \ .
\end{equation}
Firstly we must calculate $\tilde{D}^L$ which is equal to
\footnote{In the following  when we write  $Q$ inside
the commutator we mean parts of $Q$  whose commutators
with $X$ are nonzero, hence we ignore ghost contribution.
More details can be found in \cite{KlusonESFT2}.}
\begin{eqnarray}\label{Dtilde}
\tilde{D}^L_{ij}=[Q^L,P_I^L(F^I)]_{ij}
=\left[\frac{1}{\pi}\int_0^{\pi/2}d\sigma \left[
c^0(\sigma)\frac{1}{2}
((2\pi\alpha')^2P^2(\sigma)+X'^2
(\sigma))+\right.\right. \nonumber \\
+\left.\left.c^1(\sigma)(2\pi\alpha')P_K
X'^K(\sigma)\right],
 i\int_0^{\pi/2}d\sigma' P_{I}(\sigma')F^I
(\sigma')_{ij}\right]=\nonumber \\
=\frac{i}{\pi}\int_0^{\pi/2}d\sigma
\left(c^0(\sigma)\eta_{KL}X'^L(\sigma)
+ c^1(\sigma)(2\pi\alpha')
P_K(\sigma)\right)\times \nonumber \\
\times \partial_{\sigma}
\int_0^{\pi/2}d\sigma'i\delta_D(\sigma,\sigma')
\delta_I^KF^I(\sigma')_{ij}
=\nonumber \\
=-\frac{1}{\pi}\int_0^{\pi/2}d\sigma
\left(c^0(\sigma)X'^I(\sigma)\eta_{IJ}
+c^1(\sigma)(2\pi\alpha')
P_J(\sigma)\right)\partial_{\sigma}F^J(\sigma)_{ij}
=\tilde{D}^L(F(\sigma)_{ij}) \ , \nonumber \\ 
\end{eqnarray}
where $\delta_D(\sigma,\sigma')$ is delta function
with Dirichlet boundary conditions \cite{KlusonESFT2}.
In order to determine the spectrum of the fluctuations
around the new D0-brane configuration we perform field
redefinition according to our general discussion.
 As it is clear from this construction the
transformed BRST operator is the same as the 
original one
\begin{equation}
e^{\hat{K}}\left(\tilde{Q}(
e^{-\hat{K}})\right)_{ij}=Q\delta_{ij} \ 
\end{equation}
And the new field is equal to
\begin{eqnarray}\label{psitilde}
\tilde{\Psi}=(\mathcal{I}
+\hat{K})(\Psi)_{ij}=
\Psi_{ij}+\sum_k
P^L_I(\mathcal{I}F^I)_{ik}\star \Psi_{kj}-
\Psi_{ik}\star  P_I^L(\mathcal{I}F^I)_{kj}=\nonumber \\
=\Psi_{ij}-\sum_k P_I^R(F^I_{ik}\Psi_{kj})-P_I^L(\Psi_{ik}
F^I_{kj})
\ . \nonumber \\
\end{eqnarray}
Now we show that this string field carries correct winding
charge that corresponds to the string joining
i-Th and j-Th D0-brane. 
We know that the string winding charge 
is defined as
\begin{equation}
W^I=\int_0^{\pi}
d\sigma \partial_{\sigma}X^I(\sigma) \ .
\end{equation}
In order to see that this is  conserved charge
for the string obeying Dirichlet boundary
conditions  we can
calculate its commutator with the bosonic
 Hamiltonian
\begin{equation}
H=\frac{1}{4\pi\alpha'}\int_0^{\pi}
d\sigma ((2\pi\alpha')^2P^2+X'^2)
\end{equation}
and we obtain
\begin{eqnarray}
[W^I,H]=\left[
\int_0^{\pi}d\sigma X'^I(\sigma),
\frac{1}{4\pi^2\alpha'}\int_0^{\pi}
d\sigma' (2\pi\alpha')^2 P^2(\sigma')\right]\sim
\nonumber \\ \sim
\int_0^{\pi}d\sigma \int_0^{\pi}d\sigma' P^I(\sigma')
\partial_{\sigma}\delta_D(\sigma',\sigma)\sim
\int_0^{\pi}d\sigma \partial_{\sigma}\int_0^{\pi}d\sigma'
P^I(\sigma')\delta_D(\sigma',\sigma) \sim \nonumber \\
\sim \int_0^{\pi}d\sigma \partial_{\sigma} P^I(\sigma)
\sim P^I(\pi)-P^I(0)=0 \nonumber \\
\end{eqnarray}
thanks to the  Dirichlet boundary conditions on the 
boundary of the world-sheet. We also
 define 
\begin{equation}
W^I=W^I_L+W^I_R \ ,
W^I_L=\int_0^{\pi/2}
d\sigma \partial_{\sigma} X^I(\sigma) \ ,
W_R^I=\int_{\pi/2}^{\pi}
d\sigma \partial_{\sigma} X^I(\sigma) \ ,
\end{equation}
so that we have
\begin{eqnarray}\label{psitilde2}
W^I(\tilde{\Psi}_{ij})=
W^I(\Psi_{ij})-W^I(P^R_J(\sum_k
F^J_{ik}\Psi_{kj}
))-
W^I(P_J^L(\sum_k\Psi_{ik}
F_{kj}))=
  \nonumber \\
=-[W_R^I, P^R_J(F^J)](\Psi)_{ij}-
[W_L^I,P_J(F^J)](\Psi)_{ij}=\nonumber \\
=\sum_k
\left(\Psi_{ik}Y^I_{kj}-Y^I_{ik}
\Psi_{kj}\right)=
\frac{1}{\sqrt{2\pi\alpha'}}(y_i^I-y^I_j)
\Psi_{ij}
 \ , \nonumber \\
\end{eqnarray}
where we have used
\begin{eqnarray}
W^I(P_J^L(F^J\Psi))_{ij}=
W^I_L(P_J^L(F^J\Psi))_{ij}+
P_J^L(F^JW_R^I(\Psi))_{ij}=
[W^I_L,P_J^L(F^J)](\Psi)_{ij} \ ,
\nonumber \\
W^I\Psi_{ij}=(W_L^I+W_R^I)\Psi_{ij}=0
\Rightarrow W_L^I\Psi_{ij}=-W_R^I\Psi_{ij} \ ,
\nonumber \\
\end{eqnarray}
since the original configuration of D-branes corresponds
to  D-branes localized  in the common point $x^I=0$
 so that the winding charge of the strings is equal to 
zero $W\Psi_{ij}=0$.
 We also used
\begin{eqnarray}
[W^I_L,P^L_J(F^J)_{ij}]=\left[
\int_0^{\pi/2}d\sigma \partial_{\sigma}X^I(\sigma),
i\int_0^{\pi/2}
d\sigma'  P_J(\sigma')F^J(\sigma')_{ij}\right]=
\nonumber \\
=\int_0^{\pi/2}
d\sigma\partial_{\sigma}\int_0^{\pi/2}d\sigma'
\delta_D(\sigma',\sigma) \delta^I_JF^J(\sigma')_{ij}
=\nonumber \\
=\int_0^{\pi/2}
d\sigma \partial_{\sigma}F^I(\sigma)_{ij}=
(F^I(\pi/2)-F^I(0))_{ij}
=-\frac{1}{\sqrt{2\pi\alpha'}}Y^I_{ij} \ , \nonumber \\
\left[W^I_R,P^R_J(F^J)_{ij}\right]=
(F^I(\pi)-F^I(\pi/2))_{ij}=\frac{1}{\sqrt{2\pi\alpha'}}
Y^I_{ij} \ . \nonumber \\
\end{eqnarray}
The previous result suggests that the new string field
$\Psi'_{ij}$ has a winging charge proportional to the distance
between the i-Th and the j-Th D0-brane. For that reason the
ground state $\ket{ij} $ of the string connecting i-Th and j-Th
D0-brane can be represented with the operator
\begin{equation}\label{vstate}
\ket{ij}=V_{ij}=\exp \left(i\int_0^{\pi} d\sigma 
P_I(\sigma)f^I_{ij}(\sigma) \right) , \
f^I(\sigma)_{ij}=\frac{1}{\sqrt{
2\pi\alpha'}}\left(
\frac{(y_i-y_j)^I}{\pi}
\sigma -y^I_i\right) 
\end{equation}
and any  redefined 
 string field $\Psi'_{ij}$ corresponding
to the string  connecting i-Th and
j-Th D0-brane  has a form
\begin{equation}
\Psi'_{ij}=\Psi_{ij} V_{ij} \ .
\end{equation}
It is easy to see that (\ref{vstate}) has correct winding
charge. Firstly, we can calculate following commutator
\begin{eqnarray}
\left[W^I, V_{ij}\right]=
\left[W^I,\sum_{n=0}^{\infty}
\frac{1}{n!}\left(i\int_0^{\pi}d\sigma' P_J
(\sigma')f^J_{ij}(\sigma')\right)^N\right]=
\nonumber \\=
\left[\int_0^{\pi}
d\sigma \partial_{\sigma}X^I(\sigma),
i\int_0^{\pi}d\sigma' P_J(\sigma')f_{ij}^J
(\sigma') d\sigma' \right]V_{ij}=\nonumber \\
=-\int_0^{\pi}
d\sigma \partial_{\sigma}
 \int_0^{\pi}d\sigma' \delta^I_J \delta_D (\sigma',
\sigma) f^J_{ij}(\sigma')V_{ij}=
-\int_0^{\pi}
d\sigma \partial_{\sigma} f^I_{ij}(\sigma)V_{ij}=
\nonumber \\=
-
[f^I(\pi)_{ij}-f^I(0)_{ij}]V_{ij}=
\frac{1}{\sqrt{2\pi\alpha'}}(y_j-y_i)^IV_{ij} \ .
\nonumber \\ 
\end{eqnarray}
We see that $V_{ij}$ has the  correct winding
charge. Now we will calculate the action of
original BRST operator $Q$ on general new string field
$\Psi'_{ij}=V_{ij}\Psi_{ij}$. First of all, it is clear
that the action of $Q$ on the string field $\Psi$ is
the same as in the original action. For our purposes it is
important the action of $Q$ on $V_{ij}$. For that reason let us
calculate
\begin{eqnarray}\label{xV}
\left[X^I(\sigma),V_{ij}\right]=
\left[ X^I(\sigma), \sum_{N=0}^{\infty}
\frac{1}{N!} \left(i\int_0^{\pi}d\sigma' 
P_J(\sigma')f^J_{ij}(\sigma')\right)^N\right]=
\nonumber \\
=\left[ X^I(\sigma), i\int_0^{\pi}d\sigma' 
P_J(\sigma')f^J_{ij}(\sigma')\right]V_{ij}=
-\int_0^{\pi}d\sigma' \delta_J^I
\delta_D(\sigma, \sigma')f_{ij}^J(\sigma')V_{ij}=
-f_{ij}^I(\sigma)V_{ij} \  \nonumber \\
\end{eqnarray}
and consequently
\begin{eqnarray}\label{QV}
[Q,V_{ij}]=\left[
\frac{1}{\pi}\int_0^{\pi}d\sigma\left(
 \frac{1}{2}c^0(\sigma)\partial_{\sigma}
X^I(\sigma)\partial_{\sigma}X^J(\sigma)
\eta_{IJ}+\right.\right.\nonumber \\ 
\left.\left. +
c^1(\sigma)\partial_{\tau}X^I(\tau)
\partial_{\sigma}X^J(\sigma)\eta_{IJ}\right),
 V_{ij}\right]
=\nonumber \\
=\frac{1}{\pi}\int_0^{\pi} d\sigma 
\frac{(y^I_j-y^I_i)}{\sqrt{2\pi\alpha'}}V_{ij}\eta_{IJ}
\left(c^0(\sigma)\partial_{\sigma}X^J(\sigma)+
c^1(\sigma)\partial_{\tau}X^J(\sigma)\right)+\nonumber \\
+\frac{1}{\pi}
\int_0^{\pi}d\sigma 
c^0(\sigma)\frac{(y^I_j-y^I_i)\eta_{IJ}
(y^J_j-y^J_i)}{4\pi\alpha'}V_{ij} \ ,\nonumber \\
\end{eqnarray}
where we have used
\begin{eqnarray}
\left[\frac{1}{2\pi}\int_0^{\pi}
d\sigma c^0(\sigma)\partial_{\sigma}X^I(\sigma)
\partial_{\sigma}X^J(\sigma)\eta_{IJ},V_{ij}\right]=
\nonumber \\
=\frac{1}{2\pi}\int_0^{\pi}
d\sigma c^0(\sigma)\left[2\partial_{\sigma}
[X^I(\sigma),V_{ij}]\partial_{\sigma}X^J(\sigma)
\eta_{IJ}+\right. \nonumber \\
\left. +\partial_{\sigma}[X^I(\sigma),
\partial_{\sigma}[X^J(\sigma),V_{ij}]]\eta_{IJ}
\right] \ ; \\
\frac{1}{\pi}\int_0^{\pi} d\sigma c^0(\sigma)
\partial_{\sigma}
[X^I(\sigma),V_{ij}]\partial_{\sigma}X^J(\sigma)
\eta_{IJ}
=\nonumber \\
=\frac{1}{\pi}\int_0^{\pi} d\sigma c^0(\sigma)
\partial_{\sigma} (-f^I_{ij}(\sigma))V_{ij}
\partial_{\sigma}X^J(\sigma)\eta_{IJ}
=\nonumber \\
=\frac{1}{\pi}\int_0^{\pi} d\sigma c^0(\sigma)
\frac{(y^I_j-y^I_i)}{\sqrt{2\pi\alpha'}}
V_{ij}\eta_{IJ} \ ;  \\
\frac{1}{2\pi}\int_0^{\pi} d\sigma c^0(\sigma)
\partial_{\sigma}[X^I(\sigma),[X^J(\sigma),
V_{ij}]]\delta_{IJ}=\nonumber \\
=\frac{1}{2\pi}\int_0^{\pi} d\sigma c^0(\sigma)
\partial_{\sigma}[X^I(\sigma),
\frac{(y^J_j-y^J_i)}{\sqrt{2\pi\alpha'}}V_{ij}]=
\nonumber \\
=\frac{1}{2\pi}\int_0^{\pi} d\sigma c^0(\sigma)
\frac{(y^I_j-y^I_i)\eta_{IJ}
(y^J_j-y^J_i)}{4\pi\alpha'}V_{ij} \ ; \\
\left[\frac{1}{2\pi}\int_0^{\pi}
 d\sigma c^1(\sigma)\partial_{\tau}
X^I(\sigma)\partial_{\sigma}X^J(\sigma)
,V_{ij}\right]=\nonumber \\
=\frac{1}{2\pi}\int_0^{\pi}
 d\sigma c^1(\sigma)\partial_{\tau}X^I(\sigma)
\eta_{IJ}\partial_{\sigma}(
-f(\sigma)^J_{ij}V_{ij})=\nonumber \\
=\frac{1}{2\pi}\int_0^{\pi} 
d\sigma c^1(\sigma)
\partial_{\tau}X^I(\sigma)\eta_{IJ}
\frac{(y^J_j-y^J_i)}{\sqrt{2\pi\alpha'}}V_{ij} \ .
\nonumber \\
\end{eqnarray}
We see that (\ref{QV}) describes the right  action
of the BRST operator $Q$ on the string state corresponding
to the winding mode. Hence we can claim that
the solution given in (\ref{small}) correctly
describes small marginal deformation in case of
the system of $N$ D0-branes. We also see that using
of the function $F^I_{ij}$ leads to the completely
non singular solution with accord with
\cite{Takahashi3}. 
\section{Finite deformation}
As in the abelian case \cite{KlusonESFT2} we can
easily extend this approach to the case of 
of finite deformation when the solution
(\ref{solnab}) can be explicitly written as
\begin{equation}\label{nsol1}
\Phi_{0ij}=
Q(\hat{I})_{ij}+Q(\hat{K}_L(
\hat{I}))_{ij}+\sum_{n=2}^{\infty}
\frac{1}{n!}[
[Q(\hat{K}_L(\hat{I}),\hat{K}_L(\hat{I})],\hat{K}_L
(\hat{I})],\dots],]_{ij} \ , 
\end{equation}
where
\begin{equation}
[Q(\hat{K}_L(\hat{I}),\hat{K}_L(\hat{I})]_{ij}=
\sum_k Q(\hat{K}_L(\hat{I}))_{ik}\star
\hat{K}_L(\hat{I})_{kj}-
\hat{K}_L(\hat{I})_{ik}\star Q(\hat{K}_L(\hat{I}))_{kj}
\end{equation}
and commutators of higher orders are defined 
in the same manner. Now we will proceed in the
same way as in \cite{KlusonESFT2}. The first term
in (\ref{nsol1}) is given in (\ref{phis}),
(\ref{Dtilde}).  The second term
is equal to
\begin{eqnarray}\label{Uc}
\frac{1}{2}
[Q(\hat{K}_L(\hat{I}),\hat{K}_L(\hat{I})]_{ij}
=\frac{1}{2}
\sum_k Q(\hat{K}_L(\hat{I}))_{ik}\star
\hat{K}_L(\hat{I})_{kj}-
\hat{K}_L(\hat{I})_{ik}\star Q(\hat{K}_L(\hat{I}))_{kj}
=\nonumber \\
=\frac{1}{2}
\sum_k\tilde{D}^L(\mathcal{I})_{ik}\star P_J^L
(\mathcal{I}F^J)_{kj}-
P_J^L(\mathcal{I}F^J)_{ik}
\star \tilde{D}^L(\mathcal{I})_{kj}=\nonumber \\
=\frac{1}{2}\sum_kP^L_J(\tilde{D}^L
(\mathcal{I})_{ik}F^J_{kj})-
\tilde{D}^L(P_J(\mathcal{I}F^J)_{ik}F(\sigma)_{kj}) \ ,
\nonumber \\
\end{eqnarray}
where
\begin{eqnarray}
\sum_k P^L_K(\tilde{D}^L
(\mathcal{I})_{ik}F^K_{kj})=\sum_k
i\int_0^{\pi/2}d\sigma'P_K(\sigma')
\left(-\frac{1}{\pi}\int_0^{\pi/2}
d\sigma (c^0(\sigma)\partial_{\sigma}X^I(\sigma)
\eta_{IJ}+\right. \nonumber \\
+\left. c^1(\sigma)(2\pi\alpha')
P_J(\sigma))\partial_{\sigma}F^J(\sigma)_{ik}
\right)
F^K_{kj}(\sigma')=\sum_k \nonumber \\
=\left(-\frac{1}{\pi}\int_0^{\pi/2}
d\sigma (c^0(\sigma)\partial_{\sigma}X^I(\sigma)
\eta_{IJ}+c^1(\sigma)(2\pi\alpha')
P_J(\sigma))\right)\times \nonumber \\
\times i
\int_0^{\pi/2}d\sigma'P_K(\sigma')F^K(\sigma')_{ik}
\partial_{\sigma}F^J_{kj}(\sigma)-\nonumber \\
-\frac{i}{\pi}\int_0^{\pi/2}
d\sigma'\int_0^{\pi/2}d\sigma
c^0(\sigma)\partial_{\sigma}(-i\delta_D
(\sigma,\sigma'))\delta_K^I
\eta_{IJ}\partial F^J_{ik}(\sigma)
F^K_{kj}(\sigma')= \nonumber \\ 
=\sum_k\tilde{D}^L(P_K(\mathcal{I}F^K)_{ik}
F_{kj}(\sigma))
-\frac{1}{\pi}\int_0^{\pi/2}d\sigma 
c^0(\sigma)\partial_{\sigma}F^I_{ik}(\sigma)
\eta_{IJ}
\partial_{\sigma}\int_0^{\pi/2}
d\sigma' \delta_D(\sigma,\sigma')
F^J_{kj}(\sigma') \ . \nonumber \\ 
\end{eqnarray}
Then (\ref{Uc}) is equal to
\begin{equation}
\frac{1}{2}
[Q(\hat{K}_L(\hat{I}),\hat{K}_L(\hat{I})]=
-\frac{1}{2\pi}\int_0^{\pi/2}
d\sigma c^0(\sigma) \partial_{\sigma}F^I_{ik}
(\sigma)\eta_{IJ}\partial_{\sigma}F^J_{kj}(\sigma)
\end{equation} 
and according to this result  we immediately
 see that the higher order 
commutators between $\hat{K}$ in (\ref{nsol1})
are equal to  zero  and
we obtain an exact result
\begin{eqnarray}\label{nsol2}
\Phi_{0ij}=
-\frac{1}{\pi}\int_0^{\pi/2}d\sigma
\left(c^0(\sigma)X'_I(\sigma)\eta_{IJ}
+c^1(\sigma)(2\pi\alpha')
P_J(\sigma)\right)\partial_{\sigma}F^J(\sigma)_{ij}
- \nonumber \\ 
-\frac{1}{2\pi}\sum_k\int_0^{\pi/2}
d\sigma c^0(\sigma) \partial_{\sigma}F^I_{ik}
(\sigma)\eta_{IJ}\partial_{\sigma}F^J_{kj}(\sigma)
 \ . \nonumber \\
\end{eqnarray}

As in previous section we perform field
redefinition, following general
analysis given in \cite{KlusonESFT1}.
 Then the new BRST operator 
$\tilde{Q}_{ij}$ is mapped to the original one $Q\delta_{ij}$
and the new string field $\tilde{\Psi}_{ij}$
is equal to
\begin{equation}
\tilde{\Psi}_{ij}=\exp (\hat{K})(\Psi)_{ij}=
\sum_{n=0}^{\infty}\frac{1}{n!}\hat{K}^n(\Psi)_{ij} \  ,
\end{equation}
where we obtain from (\ref{psitilde})
\begin{equation}
\hat{K}(\Psi)_{ij}=
\sum_k
P^L_I(\mathcal{I}F^I)_{ik}\star \Psi_{kj}-
\Psi_{ik}\star  P_I^L(\mathcal{I}F^I)_{kj}
=-\sum_k P_I^R(F^I_{ik}\Psi_{kj})-P_I^L(\Psi_{ik}
F^I_{kj}) \ ,
\end{equation}
and from (\ref{psitilde2})
\begin{equation}
\left[W^I,\hat{K}\right](\Psi)_{ij}=(y_i^I-y^I_j)
\Psi_{ij} \ .
\end{equation}
As in the  case of the small  deformation let us work out
the action of $W$ on $\tilde{\Psi}$. We obtain
\begin{equation}
W(\tilde{\Psi})_{ij}=W(e^{\hat{K}}(\Psi))_{ij}=
[W,e^{\hat{K}}](\Psi)_{ij}=(y^I_i-y^I_j)
\tilde{\Psi}_{ij} \ ,
\end{equation}
where we have again used the fact that $W(\Psi)_{ij}=0$
and also
\begin{equation}
[W,e^{\hat{K}}]=\left[W,\sum_{N=0}^{\infty}
\frac{1}{N!}\hat{K}^N\right]=\sum_{N=0}^{\infty}
\frac{1}{N!}\hat{K}^N\left([W,\hat{K}]\right)=
e^{\hat{K}}([W,\hat{K}]) \ .
\end{equation}
It is important to stress that the upper
formula holds only on condition
$[[W,\hat{K}],\hat{K}]=0$. 

We immediately see that the new string field
$\tilde{\Psi}_{ij}$ has the  correct winding number
proportional to the distance between i-Th and j-Th
D0-branes. As in the previous section we can
write this field using the winding state $V_{ij}$
and perform the same calculation. 
As a result, the action for the fluctuation field has
the same form as the original one
\begin{equation}
S=-\frac{1}{2}\sum_{i,j}
\int \tilde{\Psi}_{ij}\star Q
(\tilde{\Psi}_{ji})-
\frac{1}{3}\sum_{i,j,k}
\int \tilde{\Psi}_{ij}\star \tilde{\Psi}_{jk}
\star \tilde{\Psi}_{ki} \ .
\end{equation}

In this section we found the  exact string
field theory solution (\ref{nsol2})
 describing marginal deformation
of the configuration of $N$ D0-branes
from the background corresponding to $N$ coincided
D0-branes to the background where D0-branes
are localized in general positions.  We believe
that our result could be helpful for other
study marginal solutions in the string field
theory.

\section{Conclusion}\label{fifth}
In this paper we have  extended our solutions of
the open bosonic string field theory that were found
 in \cite{KlusonESFT2} to the case of the open bosonic string
field theory defined on the background of $N$
D0-branes. We have shown that there exist exact
solution of the open bosonic
 string field theory that describes any
marginal deformation of the original configuration of
$N$ D0-branes localized in the origin of the space  to
the background configuration of
$N$ D0-branes in general positions. We have also seen
that using the regulation of the solution,
following \cite{Takahashi1,Takahashi2,Takahashi3}
we have obtained solutions that is completely non singular.
This fact can be used as an explanation of
the presence of the singular term in the exact solution
given in our previous work \cite{KlusonESFT2}.

The extension of this work is obvious. First of all, we would
like to perform similar analysis in the exact  
conformal field theory language, following seminal papers
\cite{SenV5,Ohmori1,Gaiotto:2002kf}. We would like to 
study other solutions using our formalism.  We hope to return
to these questions in the future. 
\\
\\
{\bf Acknowledgment}
\\
We would like to thank Ulf Danielsson and Ulf Lindstrom
for their support in our work.
This work is partly supported by EU contract 
HPRN-CT-2000-001222. This work is also supported
by the Czech Ministry of Education under Contract
No 143100006.

\end{document}